\documentclass[a4paper,12pt]{article}
\usepackage{graphicx}
\usepackage{amssymb,amsthm}
\usepackage{color}
\title{Spike-burst chimera states in an adaptive exponential integrate-and-fire neuronal network}

\begin{document}
	
\maketitle

Moises S. Santos$^1$; Paulo R. Protachevicz$^2$; Kelly C. Iarosz$^3$; Iber\^e L. Caldas$^3$; Ricardo L. Viana$^{1}$;
Fernando S. Borges$^4$; Hai-Peng Ren$^{5,6}$; Jos\'e D. Szezech Jr$^{2,7}$; Antonio M. Batista$ ^{2,7}$; 
Celso Grebogi$^{2,8}$

{\footnotesize
\noindent$^1$ Department of Physics, Federal University of Paran\'a, 80060-000,Curitiba, PR, Brazil.\\
$^2$ Graduate in Science Program - Physics, State University of Ponta Grossa, 84030-900, Ponta Grossa, PR, Brazil.\\
$^3$ Institute of Physics, University of S\~ao Paulo, 05508-900,S\~ao Paulo, SP, Brazil.\\
$^4$ Center for Mathematics Computation and Cognition, Federal University of ABC, 09606-045, S\~ao Bernardo do Campo, SP, Brazil.\\
$^5$ Shaanxi Key Laboratory of Complex System Control and Intelligent Information Processing, Xian University of Technology, Xi'an, 710048, PR China.\\
$^6$ Xian Technological University, Xi'an, 710021, PR China.\\
$^7$ Graduate in Science Program - Physics, State University of Ponta Grossa, 84030-900, Ponta Grossa, PR, Brazil.\\
$^8$ Institute for Complex Systems and Mathematical Biology, King's College, University of Aberdeen, Aberdeen, AB24 3UE, United Kingdom.
}

\section*{Abstract}
Chimera states are spatiotemporal patterns in which coherence and incoherence
coexist. We observe the coexistence of synchronous (coherent) and desynchronous
(incoherent) domains in a neuronal network. The network is composed of coupled
adaptive exponential integrate-and-fire neurons that are connected by means of
chemical synapses. In our neuronal network, the chimera states exhibit spatial
structures both with spikes and bursts activities. Furthermore, those
desynchronised domains not only have either spike or burst activity, but we
show that the structures switch between spikes and bursts as the time evolves.
Moreover, we verify the existence of multicluster chimera states.

\section{Introduction}

In dynamical systems, the word chimera has been used to describe the
coexistence of coherent and incoherent patterns \cite{omelchenko11,wildie12}.
Chimera states were first observed by Umberger et al. in 1989
\cite{umberger89}. Kuramoto and Battogtokh \cite{kuramoto02} reported
spatiotemporal patterns of coexisting coherence and incoherence in non locally
coupled phase oscillators \cite{abrams04}.

There are experimental evidences of chimera states in coupled chaotic
optoelectronic oscillators \cite{hart16}, mechanical systems
\cite{martens13,kapitaniak14}, network of electrochemical reactions
\cite{wickramasinghe14}, and populations of coupled chemical oscillators
\cite{tinsley12,nkomo13}. Totz et al. \cite{totz18} reported the existence of
spiral wave chimera states in coupled chemical oscillators. They carried out
experiments with coupled Belousov-Zhabotinsky chemical oscillators.

Numerical analysis of coupled systems have showed the coexistence of coherent
and incoherent domains \cite{omelchenko08,santos15}, and basin riddling in
chimera dynamics \cite{santos18}. Chimera sta\-tes were found in simulations of
different neuronal networks, e.g., coupled Hindmarsh-Rose neurons
\cite{hizanidis14,bera16}, coupled FitzHugh-Nagumo neurons
\cite{omelchenko15,chouzouris18}, coupled Hodgkin-Huxley neurons
\cite{sakaguchi06,glaze16}, network of integrate-and-fire neu\-rons
\cite{laing01,desmedt17}, and network composed of heterogeneous neu\-rons
\cite{laing16}. Rothkegel and Lehnertz \cite{rothkegel14} found chimera states
in small-worlds networks of excitatory integrate-and-fire-like models
\cite{rothkegel11}. Hizanidis et al. \cite{hizanidis16} observed synchronous,
metastable and chimera states in a modular organisation of the C. elegans
neuronal network. Ren et al. \cite{ren17} showed the coexistence of different
periodic states in Hindmarsh-Rose neuron network with both chemical and
electrical connections. Santos et al. \cite{santos17} reported the presence of
chimeras in the neuronal networks. They considered a network model based on the
cat cerebral cortex and identified two different chimera patterns characterised
by desynchronised spikes and bursts. In the human brain, there are analogies
between chimera state collapses and epileptic seizures \cite{andrzejak16}.

In this paper, we study a network of adaptive exponential integrate-and-fire
neurons \cite{borges17}. Brette and Gerstner \cite{brette05} introduced the
adaptive exponential integrate-and-fire (AEIF) as a simple model that mimics
the membrane potential of the neuron in vivo. Our neuronal network is a ring of
coupled AEIF, in which the neurons are connected by chemical synapses. We
observe the existence of chimera states with desynchronised spikes or bursts.
The main novelty of our work is to show chimera states whose neurons change
between spikes and bursts activities as the system evolves. In addition, we
observe multicluster chimera states that were found by Yao et al. \cite{yao15}
in Kuramoto networks of phase coupled oscillators. In our neuronal network, the
multicluster chimera states is composed not only of temporal changes between
spikes and burts, but also of domains with spike and burst patterns.

This paper is organised as follows: Section 2 introduces the neuronal network
model. In Section 3, we show and analyse the time evolution of chimera states
in our neuronal network. In the last Section, we draw our conclusions.


\section{Adaptive Exponential Integrate-and-Fire\\ Neuronal Network}

We build a network composed of $N$ coupled AEIF neurons. The neuron $i$ is
symmetrically connected with $R$ nearest neighbours on either side. The
neuronal network is composed of adaptive exponential integrate-and-fire neurons
and it is given by
\begin{eqnarray}
C_{\rm m}\frac{dV_i}{dt} & = & -g_{\rm L}(V_i-E_{\rm L})+g_{\rm L}\Delta_{\rm T}\exp 
\left(\frac{V_i-V_{\rm T}}{\Delta_{\rm T}}\right)\nonumber \\ \nonumber
& & -w_i+I_i+(V_{\rm{REV}}-V_i)\sum_{j=i-R,j\ne i}^{i+R} g_j,  \nonumber \\ 
\tau_w \frac{d w_i}{d t} & = & a (V_i-E_{\rm L})-w_i,\\ \nonumber 
\tau_{\rm s}\frac{dg_i}{dt} & = & -g_i,
\label{eqIFrede}
\end{eqnarray}
where $V_i$ is the membrane potential, $w_i$ is the adaptation current, and
$g_{i}$ is the synaptic conductance. We consider: membrane capacitance
$C_{\rm m}=200$ pF, resting potential $E_{\rm L}=-70$ mV, leak conductance
$g_{\rm L}=12$ nS, slope factor $\Delta_{\rm T}=2$ mV, spike threshold potential
$V_{\rm T}=-50$ mV, adaptation time constant $\tau_w=300$ ms, level of
subthreshold adaptation $a=2$ nS, synaptic time constant $\tau_{\rm s}=2.728$
ms, injection of current $I_i=500$ pA, and synaptic reversal potential
$V_{\rm REV}=0$ mV (excitatory synapses). When $V_i$ is larger than a threshold,
$V_i>V_{\rm thres}$ \cite{naud08}, $V_i$, $w_i$, and $g_i$ are updated following
the rules
\begin{eqnarray}
V_i &\to & V_{\rm r}, \nonumber\\
w_i &\to& w_i + b,\\
g_{i} &\to& g_{i}+g_{\rm ex}, \nonumber
\end{eqnarray}
where $V_{\rm r}=-58$ mV, $b=70$ pA, and $g_{\rm exc}$ is the intensity of the
excitatory synaptic conductance.

\begin{figure}[hbt]
\centering
\includegraphics[scale=0.4]{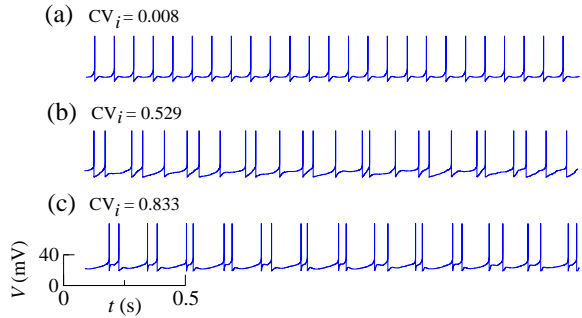}
\caption{Action potential of the neuron $i$ in our neuronal network for (a)
spikes (${\rm CV}_i=0.008$), (b) spikes and bursts (${\rm CV}_i=0.529$), and
(c) bursts (${\rm CV}_i=0.833$).} 
\label{fig1}
\end{figure}

Depending on the control parameters, the AEIF neurons can generate spike or
burst firings, and the network can also exhibit synchronous and desynchronous
behaviour. In all simulations, we consider that the individual uncoupled
neurons perform spike activities for the chosen parameters. With regard to the
initial conditions, $V_i$ and $w_i$ are randomly distributed in the intervals
$[-58,-43]$mV and $[0,70]$pA, respectively. We analyse the solution of the
neuronal network model during $2$s and discard a transient time equal to $4$s.

To identify spike or burst activities, we calculate the coefficient of
variation (${\rm CV}_i$)
\begin{equation}
{\rm CV}_i = \frac{\sigma_{{\rm ISI}_i}}{{\overline{\rm ISI}_i}},
\end{equation}
where $\overline{\rm ISI}_i$ is the mean value of the time difference between
two consecutive firings (inter-spike interval) of the neuron $i$ and
$\sigma_{{\rm ISI}_i}$ is the standard deviation of ${\rm ISI}_i$. In Fig.
\ref{fig1}, we see the action potential of the neuron $i$ for (a) spikes
(${\rm CV}_i=0.008$), (b) spikes and bursts (${\rm CV}_i=0.529$), and (c)
bursts (${\rm CV}_i=0.833$).

\begin{figure}[hbt]
\centering
\includegraphics[scale=0.6]{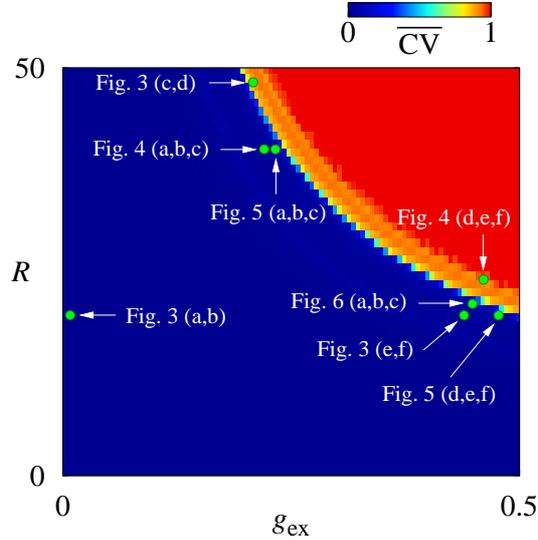}
\caption{(Colour online) Parameter space $R\times g_{\rm ex}$ for
$\overline{\rm CV}$, where we consider $N=1000$ coupled AEIF neurons and
$50$ different initial conditions. The regions have spiking
($\overline{\rm CV}<0.5$), bursting ($\overline{\rm CV}\geq 0.5$) neurons,
and the coexistence of bursting and spiking can be seen according to the
colour.} 
\label{fig2}
\end{figure}

The mean value of ${\rm CV}$ ($\overline{\rm CV}$) is given by
\begin{equation}
\overline{\rm CV}=\frac {1}{N}\sum_{i=1}^{N} {\rm CV}_i.
\end{equation}
For $\overline{\rm CV}<0.5$ and $\overline{\rm CV}\geq 0.5$ the neuronal
network exhibits spikes and bursts, respectively \cite{protachevicz18}. Figure
\ref{fig2} shows the parameter space $R\times g_{\rm ex}$ for $\overline{\rm CV}$
in colour scale. The regions for $\overline{\rm CV}<0.5$ and
$\overline{\rm CV}\geq 0.5$ correspond to spike and burst activities,
respectively. In the transition region, we identify the coexistence of spike and
burst behaviours for different neurons in the network. Synchronous and
desynchronous behaviours also occur for different values as $R$ and $g_{\rm ex}$.


\begin{figure}[hbt]
\centering
\includegraphics[scale=1]{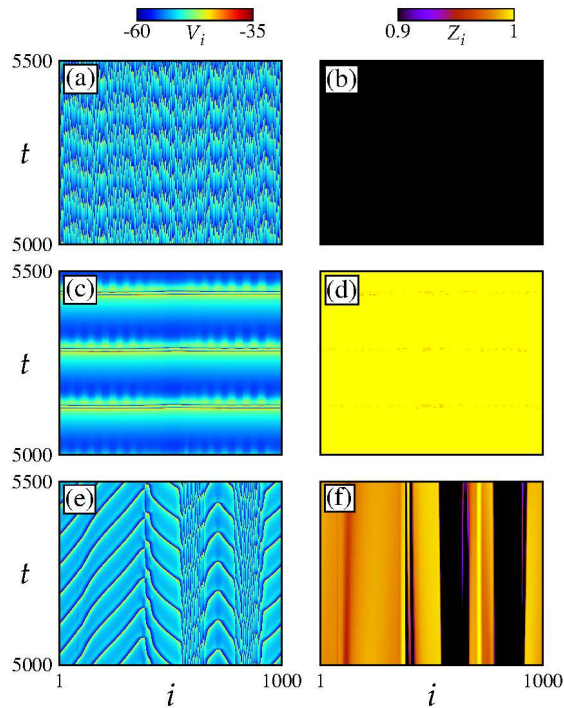}
\caption{(Colour online) Time evolution of $V_i$ and $Z_i$ of each neuron $i$
for incoherent pattern for $R=20$ and $g_{\rm ex}=0.01$ nS ((a) and (b)),
synchronised behaviour for $R=48$ and $g_{\rm ex}=0.21$ nS ((c) and (d)), and
chimera state for $R=20$ and $g_{\rm ex}=0.44$ nS ((e) and (f)).} 
\label{fig3}
\end{figure}

\section{Chimera States}

In the chimera states there are spatiotemporal patterns characterised by the
coexistence of coherent and incoherent domains. The spatial coherence and
incoherence can be identified by the local order parameter \cite{kuramoto84}
\begin{equation}
Z_j(t)=\left|\frac{1}{2\delta +1}\sum_{|j-k|\le\delta}
       {\rm e}^{{\rm i}\phi_k(t)}\right|, \hspace{1cm} k = 1, ..., {N}.
\end{equation}
The phase is defined as
\begin{equation}
\phi_k(t)=2\pi m+2\pi\frac{t-t_{k,m}}{t_{k,m+1}-t_{k,m}},
\end{equation}
where $t_{k,m}$ is the time of the $m$th spike of the neuron $k$,
$t_{k,m}<t<t_{k,m+1}$, and the spike happens for $V_k>V_{\rm thres}$. In our
simulations, we use $\delta=5$ and consider a pattern to be synchronised when
$Z_j(t)>0.9$. The coherent (synchronised) and incoherent (desynchronised)
domains are identified with a minimum size equal to $2\delta+1$ neighbours.

In Fig. \ref{fig3}, we present different types of dynamic behaviour by varying
the parameters $R$ and $g_{\rm ex}$. Figures \ref{fig3}(a) and \ref{fig3}(b)
display the neuronal network having incoherent pattern. Depending on the
parameters, the network can exhibit synchronisation (Figs. \ref{fig3}(c) and
\ref{fig3}(d)). We find chimera states with coexisting synchronous and
desynchronous domains, as shown in Figs. \ref{fig3}(e) and \ref{fig3}(f). 

\begin{figure}[hbt]
\centering
\includegraphics[scale=0.95]{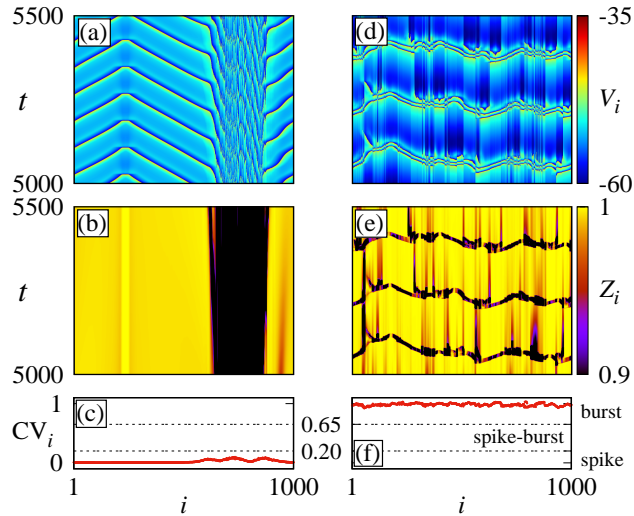}
\caption{(Colour online) $V_i$ and $Z_i$ in colour scale for $t\times i$, and
${\rm CV}_i\times i$. Chimera with spiking neurons for $R=40$ and
$g_{\rm ex}=0.22$ nS ((a), (b), and (c)), and chimera with bursting neurons for
$R=24$ and $g_{\rm ex}=0.46372$ nS ((d), (e), and (f)).}
\label{fig4}
\end{figure}

Our AEIF neuronal network can exhibit different firing patterns, such as
spikes, bursts, or both spikes and bursts. Figure \ref{fig4} exhibits the
values of $V_i$, $Z_i$, and ${\rm CV}_i$ for chimera with spiking neurons
(\ref{fig4}(a), \ref{fig4}(b), and \ref{fig4}(c)) and chimera with bursting
neurons (\ref{fig4}(d), \ref{fig4}(e), and \ref{fig4}(f)). 

\begin{figure}[hbt]
\centering
\includegraphics[scale=0.9]{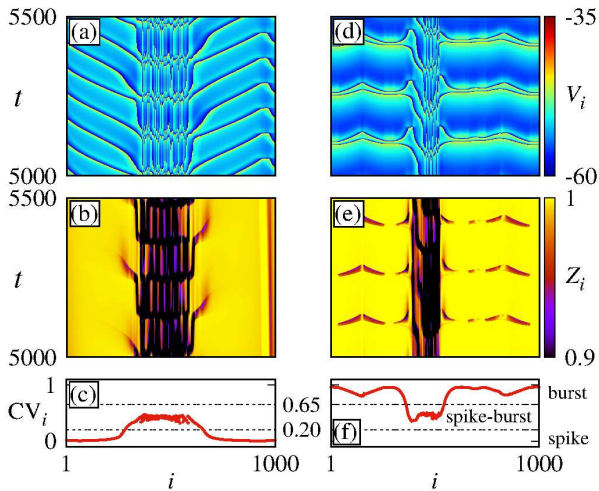}
\caption{(Colour online) $V_i$ and $Z_i$ in colour scale for $t\times i$, and
${\rm CV}_i\times i$. The SBC with synchronised spikes for $R=40$ and
$g_{\rm ex}=0.233$ nS ((a), (b), and (c)), and synchronised bursts for $R=20$ and
$g_{\rm ex}=0.48$ nS ((d), (e), and (f)).}
\label{fig5}
\end{figure}

A phenomenon not seen before is that the chimera states keep switching, in the
desynchronous domains, between spikes and bursts as the system evolves in time,
which we call spike-burst chimera (SBC). The SBC is found through $Z_j(t)$ and
${\rm CV}_i$. We identify SBC when the ${\rm CV}_i$ values are in the interval
$[0.2, 0.65]$. In Fig. \ref{fig5}, we calculate $V_i$, $Z_i$, and ${\rm CV}_i$
for parameters when the SBC states are present. Figs. \ref{fig5}(a),
\ref{fig5}(b), and \ref{fig5}(c) display a SBC with synchronised spikes for
$R=40$ and $g_{\rm ex}=0.233$ nS. The SBC can have synchronised bursts, as shown
in Figs. \ref{fig5}(d), \ref{fig5}(e), and \ref{fig5}(f) for $R=20$ and
$g_{\rm ex}=0.48$ nS.

\begin{figure}[hbt]
\centering
\includegraphics[scale=1.0]{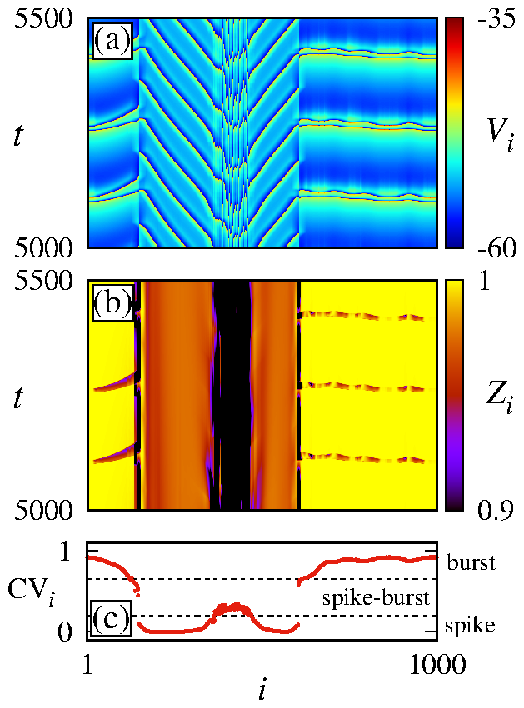}
\caption{(Colour online)(a) $V_i$ and (b) $Z_i$ in colour scale for
$t\times i$, and (c) ${\rm CV}_i\times i$. Figure exhibits multicluster chimera
states, where we consider $R=21$ and $g_{\rm ex}=0.45$ nS.}
\label{fig6}
\end{figure}

In Fig. \ref{fig6}, we show the presence not only SBC, but also multicluster
chimera states \cite{yao15}. The network has groups of neurons with different
patterns, such as spikes (${\rm CV}_i\leq 0.20$), bursts
${\rm CV}_i\geq 0.65$), and a mixture of spikes and bursts
($0.20<{\rm CV}_i<0.65$). Beurrier et al. \cite{beurrier99} reported that the
transition from spike to mixed burst activities in subthalamic nucleus neurons
of rat and primates is one of the features of Parkinson's disease. There are
two groups with spiking neurons, where one group has $217$ neurons and the
other $134$ neurons, as well as one group with $511$ bursting neurons. The SBC
is identified by means of three groups with $22$, $105$, and $11$ neurons that
change between spike and burst patterns over time. 

\begin{figure}[hbt]
\centering
\includegraphics[scale=0.4]{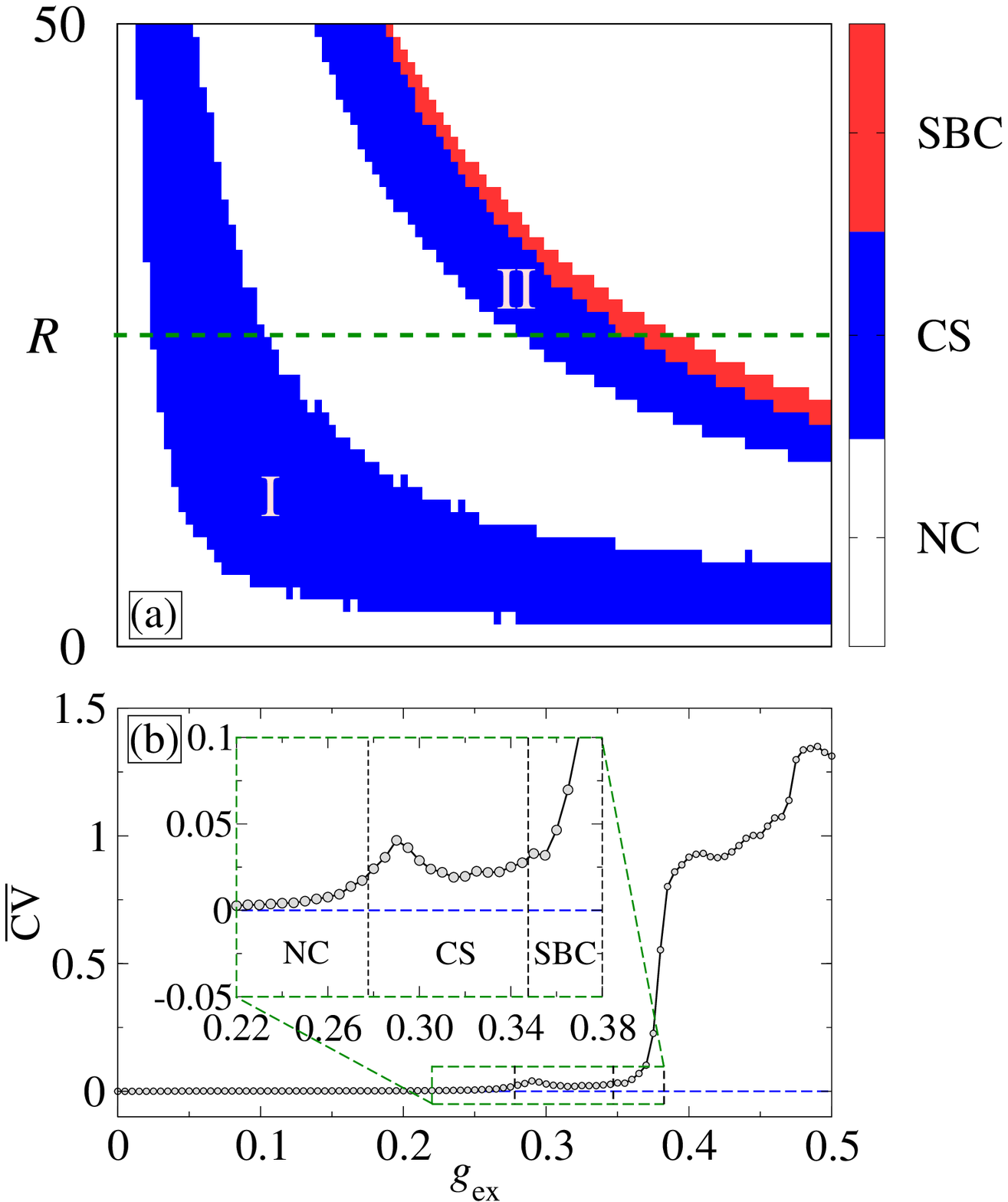}
\caption{(Colour online) (a) Parameter space $R\times g_{\rm ex}$ showing regions
with chimera (CS) and spike-burst chimera (SBC), as well as without chimera
(NC), where we consider $N=1000$ coupled AEIF neurons and $50$ different random
initial conditions. (b) $\overline{\rm CV}$ as a function of $g_{\rm ex}$ for
$R=25$ (green dashed line in Fig. \ref{fig7}(a)).} 
\label{fig7}
\end{figure}

Figure \ref{fig7}(a) shows in the parameter space $R\times g_{\rm ex}$ the regions
for chimera state (CS) and spike-burst chimera (SBC) states, as well as the
regions where there are no chimera states (NC). We perform an average of $50$
different random initial conditions to compute each point in the parameter
space. The regions in blue and white colours represent CS and NC, respectively.
The small region in red colour denotes the values of $R$ and $g_{\rm ex}$ in
which SBC is observed. Figure \ref{fig7}(b) displays the values of
$\overline{\rm CV}$ as a function of $g_{\rm ex}$ for $R=25$. We observe that
$\overline{\rm CV}$ increases in the transitions from NC to CS and from CS to
SBC, showing a scenario from spike to burst activities. When we fix $g_{\rm ex}$
and vary $R$, we find a similar behaviour. There are two disjoint regions with
chimera states (CS) in Fig. \ref{fig7}(a), where the region II has a standard
deviation of ISI greater than the region I, consequently $\overline{\rm CV}$ in
the region II is greater than in the region I, as shown in Fig. \ref{fig7}(b).
 
\section{Conclusions}

We study an adaptive exponential integrate-and-fire neuronal network. In the
network, each neuron is symmetrically coupled to the nearest neighbours. The
connectivity between the neurons is given by excitatory synapses. Depending on
the control parameters, the neurons can exhibit spike or burst activities.

Researches have reported the coexistence of spatiotemporal patterns, known as
chimera states. There are evidences of chimera states in the brain, e.g.,
unihemispheric slow-wave sleep in some mammals. The coexistence of synchronous
and desynchronous domains has been observed in neuronal network models.

In our network, chimera states are found by varying the number of nearest
neighbours and the excitatory synaptic conductance. Depending on the coupling
strength, multichimera state can arise for small $R$ values
\cite{omelchenko13}. We verify the existence of different types of chimera
states according to the spike and burst patterns. In this work, we show the
existence of chimera states with neurons that change between spike and burst
activities as the system evolves in time. Moreover, we also identify
multicluster chimera states composed of different groups of neurons with spike
and burst patterns, as well as spikes and bursts changing over time.


\section*{Aknowledgments}
We wish to acknowledge the support: Funda\c c\~ao Arau\-c\'aria, CNPq\\
$(150701/2018-7)$, CAPES, and FAPESP (2015/07311-7 and 2018/03211-6).

\end{document}